\documentclass[iop]{emulateapj}

\newcommand{\kms}{~km~s$^{-1}$}
\newcommand{\htwo}{H$_2$}
\newcommand{\percc}{~cm$^{-3}$}

\slugcomment{Accepted for publication in Ap.J.}

\shorttitle{Zpectrometer/GBT detections}
\shortauthors{Harris et al.}

\begin{document}

%%%%%%%%%%%%%%%%%%%%%%%%%%%%%
%% To Do %%%%%%%%%%%%%%%%%%%%
%%%%%%%%%%%%%%%%%%%%%%%%%%%%%

\title{CO $J=1-0$ spectroscopy of four submillimeter galaxies with the
Zpectrometer on the Green Bank Telescope}

\author{A.I.~Harris\altaffilmark{1}, A.J.~Baker\altaffilmark{2},
  S.G.~Zonak\altaffilmark{1}, C.E.~Sharon\altaffilmark{2},
  R.~Genzel\altaffilmark{3,4},
  K.~Rauch\altaffilmark{1},
  G.~Watts\altaffilmark{5}, R.~Creager\altaffilmark{5}}

\altaffiltext{1}{Department of Astronomy, University of Maryland,
  College Park, MD 20742; \emph{harris, szonak, rauch @astro.umd.edu}}
\altaffiltext{2}{Department of Physics and Astronomy, Rutgers, the
  State University of New Jersey, Piscataway, NJ, 08854-8019; \emph{ajbaker,
    csharon @physics.rutgers.edu}}
\altaffiltext{3}{Max-Planck-Institut f\"ur extraterrestrische Physik,
  Giessenbachstrasse~1, D-85741 Garching, Germany; \emph{genzel@mpe.mpg.de}}
\altaffiltext{4}{Department of Physics, Le Conte Hall, University of
  California, Berkeley, CA 94720}
\altaffiltext{5}{National Radio Astronomy Observatory, P.O. Box 2,
  Green Bank, WV 24944; \emph{gwatts, rcreager @nrao.edu}}

\begin{abstract}
  We report detections of three $z \sim 2.5$ submillimeter-selected
  galaxies (SMGs; SMM~J14011+0252, SMM~J14009+0252, SMM~J04431+0210)
  in the lowest rotational transition of the carbon monoxide molecule
  (CO $J = 1-0$) and one nondetection (SMM~J04433+0210).  For the
  three galaxies we detected, we find a line-integrated brightness
  temperature ratio of the $J = 3-2$ and $1-0$ lines of $0.68 \pm
  0.08$; the $1-0$ line is stronger than predicted by the frequent
  assumption of equal brightnesses in the two lines and by most
  single-component models.  The observed ratio suggests that mass
  estimates for SMGs based on $J = 3-2$ observations and $J = 1-0$
  column density or mass conversion factors are low by a factor of
  1.5.  Comparison of the $1-0$ line intensities with intensities of
  higher-$J$ transitions indicates that single-component models for
  the interstellar media in SMGs are incomplete.  The small dispersion
  in the ratio, along with published detections of CO lines with
  $J_{upper} > 3$ in most of the sources, indicates that the emission
  is from multi-component interstellar media with physical
  structures common to many classes of galaxies.  This result tends to
  rule out the lowest scaling factors between CO luminosity and
  molecular gas mass, and further increases molecular mass estimates
  calibrated against observations of galaxies in the local universe.
  We also describe and demonstrate a statistically sound method for
  finding weak lines in broadband spectra that will find application
  in searches for molecular lines from sources at unknown redshifts.
\end{abstract}


\section{Introduction}\label{sec:intro}

Observations and models of the extragalactic
far-infrared/submillimeter background
\citep[e.g.,][]{puge96,fixs98,laga03} indicate that a large fraction
of cosmic star formation has taken place behind a veil of dust. At
high redshift, this conclusion is consistently affirmed by comparisons
of obscured and unobscured star formation in optically selected galaxy
populations \citep[e.g.,][]{redd08}.  A more striking pattern
emerges when one considers {\it all} galaxy populations, however: the systems
forming stars at the highest rates (i.e., having the highest
bolometric luminosities) are also the dustiest \citep{adel00}.  This
trend reaches an extreme in the case of submillimeter galaxies (SMGs),
first identified over a decade ago as bright ($> 5$~mJy) sources in
$850$~$\mu$m surveys with the Submillimeter Common-user Bolometer
Array (SCUBA) on the James Clerk Maxwell Telescope (JCMT)
\citep[e.g.,][]{smai97,barg98,hugh98}.  SMGs have faint X-ray
counterparts \citep{alex05} and show the disturbed morphologies
characteristic of major mergers \citetext{\citealp{cons03}, cf.\
  \citealp{dave10}}, suggesting they are important sites of mass
assembly as well as star formation.  However, their high obscuration
has also proved challenging for detailed study, including determining
their precise redshifts and masses, and understanding their
interstellar media and star formation processes.  While photometric
techniques shed some light on SMGs' redshift distribution
\citep[e.g.,][]{aret03}, obtaining more than a handful of
spectroscopic redshifts for SMGs and their warmer analogs has required
a painstaking effort of radio continuum mapping followed by optical
spectroscopy \citep{chap03,chap04,chap05}, whose results have been
securely confirmed by the detection of CO emission lines at millimeter
wavelengths \citep{neri03,greve05,tacconi06,chapman08,bothwell10}.

Redshift determination is the first step towards addressing the
crucial question of SMGs' masses.  Translation of SMGs' angular
clustering strength \citep{blai04, weiss09a} to a linear correlation
length and dark matter halo mass depends sensitively on their exact
redshift distribution, although a large characteristic halo mass can
be independently estimated from cosmological simulations assuming a
proportionality between dark matter accretion and star formation rates
\citep{gene08}.  The total dynamical masses of the galaxies themselves
are also large, as first suggested by measurements of large CO line
widths \citep[e.g.,][]{fray98,frayer99} and confirmed by spatially
resolved CO mapping
\citep[e.g.][]{downes03,genzel03,tacconi06,tacconi08, bothwell10}.
However, given the challenges of obtaining such mapping and the
complications of high extinction for stellar mass determinations
\citep[e.g.,][]{hain10}, estimating SMGs' molecular gas masses from
their CO line luminosities remains a useful way to place lower limits
on their total masses.  Accurate molecular gas masses are also
required to determine SMGs' gas mass fractions and star formation
efficiencies, important inputs for understanding their evolutionary
status and the likely properties of their descendants
\citep{baug05,swin09}.

In this paper, we report observations of four SMGs, drawn from the
SCUBA Lens Survey (SLS) sample of \citet{smail02}, with the
ultrawide-bandwidth Zpectrometer cross-correlation spectrometer on the
100~m diameter Robert C. Byrd Green Bank Telescope (GBT).  For the
three of our targets that are brightest at 850~$\mu$m, the
Zpectrometer has successfully detected the lowest ($J = 1-0$)
rotational transition of carbon monoxide (CO), and we undertake a
joint analysis with published CO $J = 3-2$ detections for this trio of
massive SMGs at $z \sim 2.5$.  

Studies of our own and nearby galaxies have established the CO $J = 1
- 0$ line ($\nu_{\rm rest} = 115.27~{\rm GHz}$) as a proxy for tracing
the molecular hydrogen (\htwo) that forms giant molecular clouds.  CO
is the most abundant molecule after \htwo\ and, unlike \htwo, has a
permanent dipole moment that allows it to radiate efficiently.  The
magnitude of the molecule's dipole moment and close spacing between
the lowest rungs of its rotational ladder allow it to be collisionally
excited to trace gas particle densities above a few hundred per cubic
centimeter and temperatures above 5~K.  The strong C--O bond helps
keep the molecule stable against dissociation by ultraviolet light and
shocks, so CO probes active as well as quiescent regions.  Transitions
between rotational states with increasing rotational quantum number
$J$ trace increasingly warm and dense gas. Rest-frame submillimeter
lines trace particle densities of $\sim 10^{4-5}$\percc\ and kinetic
temperatures of several tens to a few hundred K.

For all its advantages as a molecular gas tracer, observations of the
$J = 1-0$ transition at high redshift have been hampered by its
relatively low observed frequency.  Observations of the line are
rapidly becoming easier as technology improves, as the run of
publication dates for detections of galaxies at $z > 1$ shows
\citep{papadopoulos01, greve03, klamer05, hainline06, riechers06,
  swinbank10, carilli10, ivison10, aravena10, negrello10, frayer10,
  ivison10a}.  Peak flux density from optically thick and thermalized
emission scales roughly as frequency squared, so the $1-0$ lines from
distant galaxies are expected to be an order of magnitude weaker than
the $3-2$ and other mid-$J$ lines from the same sources.  The 100
meter diameter Green Bank Telescope combines an enormous collecting
area and a low centimeter-wave system temperature to approach the
necessary sensitivity level.  Stability is a key consideration for
long integrations, and motivated our construction of the Zpectrometer,
a cross-correlation spectrometer optimized for line searches that
instantaneously covers the GBT Ka-band receiver's 25.6 to 36~GHz band.
In this band the CO $J = 1-0$ line redshifts over a $2.2 \leq z \leq
3.5$ range that includes the peak of the \citet{chap05} SMG redshift
distribution.  The Zpectrometer and correlation receiver architecture
have improved stability to the point that detection of $1-0$ emission
is relatively straightforward for sources with extreme (${\cal M}
\gtrsim 10$) magnifications due to gravitational lensing
\citep{swinbank10, negrello10, frayer10}.  In this paper we discuss
observations of weaker CO $1-0$ lines from galaxies that have
considerably less magnification than the brightest sources.

The prime goal of our observations was to measure the ratios of $1-0$
to higher-$J$ line fluxes to explore the physical conditions
in our targets' molecular gas and to test standard assumptions in the
use of empirical conversion factors to relate their CO luminosities to
their molecular gas masses.  CO $J = 1-0$ spectra of these galaxies
are especially important because they constrain the state of their
molecular interstellar media and may reveal the presence of massive
reservoirs of extended cool gas that do not appear in lines from more
excited states.  In addition to tracing cool gas, the $1-0$ line is
essential for interpreting mid-$J$ lines to constrain the properties
of warmer gas.  In the local universe, multi-line observations of
nearby starburst and active galactic nuclei typically show low- and
high-excitation gas components \citetext{e.g., \citealp{wild92, guesten93,
    mao00, ward03, greve09}}, with the two components distinct in low-
and high-$J$ lines but contributing jointly to mid-$J$ lines.  Fluxes
in the $1-0$ and $2-1$ lines are essential for characterizing the cool
component well enough to determine the fraction of the mid-$J$
emission that comes from each component.

A secondary outcome of our observations was to test the use of the
Zpectrometer's large fractional bandwidth ($\Delta f/ f_{\rm mean} =
34$\%) for blind redshift searches toward targets identified in
continuum surveys.  The Zpectrometer's instantaneous redshift coverage
is five to ten times larger than those provided by current
millimeter-wave interferometer bandwidths, offering the promise of
quick CO redshift determinations for SMGs without waiting for radio
continuum mapping or optical spectroscopy.  To that end, we developed
a statistical test for line detection appropriate for long
observations across wide bandwidths.  The ability to identify sources
directly from continuum positions known to within the
$22^{\prime\prime}$ size of the GBT's Ka-band beam eliminates some
selection effects that bias detection towards excited molecular gas.

This paper has three further sections and an appendix.
Section~\ref{sec:obs} describes our observational technique and
instrument. Section~\ref{sec:res} covers our basic results for each
source, including an overview of a detection method suitable for
wideband spectroscopy.  Section~\ref{sec:disc} contains discussion of
the physical conditions in the galaxies we detected, the implications,
and a brief summary. An appendix describes the details of the
detection statistic whose use we demonstrate here.

\section{Observations}\label{sec:obs}

The combination of the Zpectrometer cross-correlation spectrometer and
dual-channel Ka-band correlation receiver is specifically intended for
wideband observations of weak lines.  The system's use of correlation
makes it the single-dish equivalent of a two-element spatial
interferometer \citep{blum59, harris05}, with a similar promise for
high stability.  The spectrometer has moderate velocity resolution
that is well matched to extragalactic observations: its spectral
response to a monochromatic line is a sinc($x$) function with 20~MHz
full width at half maximum, corresponding to about 190\kms\ at band
center.  A correlation spectrometer is immune to some systematic
effects in line detection experiments, as the narrow spectral feature
corresponding to a line is not produced by a single detector, but is
created by coherent structure across hundreds of independent lags.

Physically, the Zpectrometer analog lag cross-correlator is installed
next to the Ka-band receiver on the National Radio Astronomy
Observatory's 100 meter diameter Green Bank Telescope \citep{zp07}.
Even in a digital age, analog multiplication still retains advantages
of low power dissipation and low radiated emissions for broadband
spectroscopy.  Transistor analog multipliers separated by transmission
line delays cascade to form the Zpectrometer's lag correlators.  Sets
of 256 lags are packaged in four identical independent
cross-correlator units, each with 3.5~GHz bandwidth.  A four-channel
downconverter splits the receiver's IF band to stack the correlators
in frequency space.  Receiver performance at high frequency limits the
bandwidth, so the spectra in this paper cover 10.5~GHz, the three
lowest-frequency correlator sub-bands.  \citet{harris05} covers the
details of how the combination of a correlation receiver front end and
analog lag cross-correlator backend differences the power between the
receiver's two input feed horns.  This electronic differencing greatly
reduces the effects of amplifier $1/f$ gain noise, the dominant source
of instability in good total power radiometers.  Layers of optical
switching (chopping, wobbling, nodding, beamswitching,
double-beamswitching, etc.; the terminology is mixed) and electronic
phase switching remove other nonideal signals to a very high degree,
leaving spectra that are very clean compared with a conventional total
power systems.

While the correlation architecture greatly improved stability,
switching the beam position on the sky by moving the telescope
subreflector (chopping) at a 10 second period was still necessary for
usable stability.  A chop throw of 78~arcsec, equal to the angular
separation of the correlation receiver's two feed horns, alternately
placed the source on one of the two horns, optically switching the
source between the receiver's ``plus'' and ``minus'' sense beams.
Differencing the spectra from the two subreflector positions (tilts)
reduced electronic imbalances, but the slight difference in telescope
illumination in the two subreflector positions introduced spectral
structure from optical imbalance with a peak-peak amplitude of
approximately 60~mJy.  Changing the antenna position to view a
reference position on the nearby sky and then differencing these
``source'' and ``reference'' subreflector-switched spectra removed the
optical imbalance to a high degree.  This is the classical
double-beamswitching or chop-and-nod pattern common to
short-wavelength radio and infrared astronomy; we used an 8 minute
cycle time and telescope moves of no more than a few degrees.  Rather
than spending half of the time observing blank sky, we took advantage
of the system's fixed-tuned operation and wide bandwidth by switching
between two sources close in sky positions.  This increases observing
efficiency by a factor of two at the cost of not preserving individual
source continuum levels, and risking that lines from the two sources
will fall at the same frequencies and cancel to a greater or lesser
extent.  We felt the gain in time offset the risk, and we preserved
the data from each chop side to recover from this eventuality.
Residual imbalance in the receiver caused some fluctuating large scale
structure and an occasionally strong ripple across the spectral
baseline on timescales short compared with the optical switching
times.  While this structure tended to cancel after long integrations,
it was still present in the spectrum and added nonideal noise.
Although the noise amplitude in spectra decreased as the square root
of integration time, excess noise in the receiver was a factor of two
to four higher than the radiometer equation predicts.

We made gain corrections across the band and established the intensity
scale by dividing the source minus reference spectra by spectra of the
standard radio flux calibrators obtained in the same observing
sessions.  We took absolute fluxes of 0.77 and 1.9~Jy at 32.0~GHz for
3C48 and 3C286 from The \citeauthor{AstAlm2008}.  Observations of
Mars, compared against a physical model for Mars' emission
\citetext{B.~Butler, priv.\ comm.}, independently verified the flux
scale within a few percent and established the quasars' spectral
indices across the band.  We tracked the total power at the
Zpectrometer's four-channel downconverter to monitor the system
temperature, but system temperature measurements were complicated by
the presence of nonideal phase noise, which does not appear in total
power.

Stepping monochromatic signals with 8~MHz spacings across the receiver
band calibrated the system phase and established the correlator's
spectral frequency scale \citep{har_zmu01, harris05}.  Signals from
one of the GBT's microwave synthesizers were injected between one of
the feed horns and the receiver's input hybrid for this purpose.  To
simplify this phase calibration, the system local oscillators run at
fixed frequencies, so the system frequency scale is topocentric.
Doppler shift corrections to bring the velocity scale to a local
standard of rest (LSR) scale were applied in data reduction.  We used
version D of the Zpectrometer's data reduction pipeline, written in
GBTIDL \citep{GBTIDLref}, to make calibrated spectra and quick-look
evaluation during observations. Further data analysis was in R
language \citep{Rref} routines.

The telescope pointing was generally excellent when the wind was low
($\leq 3$~m~s$^{-1}$) and for elevations between 75\degr and 20\degr.
We pointed on a compact source near the astronomical targets once per
hour, finding typical corrections of 0.1 arcmin or below, or about a
third of the 0.35 arcmin beamwidth (0.84 of the peak amplitude for a
Gaussian beam) at 30~GHz.

We took spectra of the pointing source hourly to monitored systematic
gain changes, and periodically monitored receiver gain with a
modulated noise source at its input.  Optical gain changes from
pointing and focus errors were dominant, and we found that the
systematic overall calibration drifted by a maximum of 20\% over an
hour, less at lower frequencies than high.  The drifts produced a bias
that could slightly underpredict CO $J = 1-0$ line strengths.

\section{Results}\label{sec:res}

\begin{deluxetable*}{lccrr}
\tabletypesize{\scriptsize}
%\rotate
\tablecaption{Summary of observations.  A horizontal line separates the
  pairs of sources that were observed together. The total integration
  time $t_{int}$ is the elapsed observing time for each of the
  difference spectra, Figs.~\ref{fig:mamo} and \ref{fig:fred}.
\label{tab:summ}}
\tablewidth{0pt}
\tablehead{
 \colhead{Name} &
 \colhead{RA (J2000)} &
 \colhead{Dec (J2000)} &
 \colhead{No.\ sess.} &
 \colhead{$t_{int}$}
}
\startdata
SMM~J04431+0210 & 04:43:07.10 & $+$02:10:25.1 & 9 & 9.8\\ % Fred
SMM~J04433+0210 & 04:43:15.00 & $+$02:10:02.0 &  &  \\ % Ginger
\hline
SMM~J14011+0252 & 14:01:04.96 & $+$02:52:23.5 & 3 & 3.1\\ % Max
SMM~J14009+0252 & 14:00:57.68 & $+$02:52:48.6 &  & \\ % Moritz
\enddata
\end{deluxetable*}

We observed the sources in pairs to cancel optical offsets, as
described in \S\ref{sec:obs}, so we summarize the results by source
pair.  Positions, numbers of observing sessions, and total integration
times $t_{int}$ for each pair are listed in Table~\ref{tab:summ}.

Our data analysis for the spectra in this paper is untuned: we used an
unweighted average of all the data from all of the sessions for all
sources.  Time-varying nonideal noise dominated the spectral
structure, so weighting by system temperature was not appropriate, and
ad hoc weighting or editing based on measured session fluctuations or
structure in individual spectra was not justified.  No baseline
structure has been removed other than narrowband filtering in the
Fourier domain to remove an approximately $300$~MHz period ripple
produced in the receiver.  Offsets from zero flux density in the
spectra are reproducible but are different for different source pairs:
they give the continuum level differences between the sources in the
pairs.  There is no sign of discontinuity between the three correlator
sub-bands, even on strong continuum sources, a sign that the
correlator's response is very linear in power.

\paragraph{SMM~J14011+0252 and SMM~J14009+0252}

\begin{figure*}
%\epsscale{.80}
\vspace{2.5cm}
\plotone{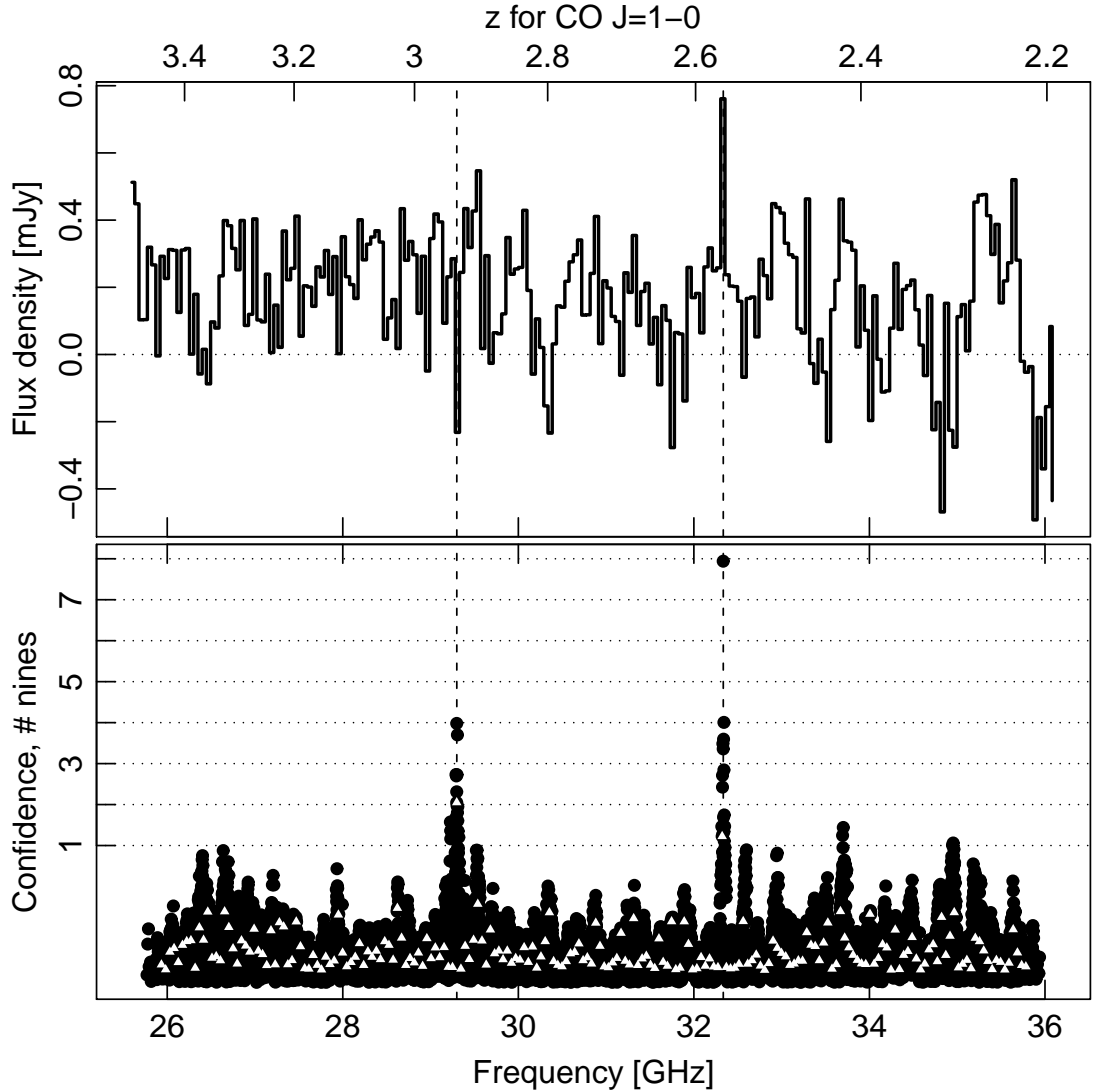} %MaMoPlot.eps}
\caption{Difference spectrum of SMM~J14011+0252 and SMM~J14009+0252
  (upper panel) and confidence plot (lower panel).  The triangles in
  the confidence plot mark the points corresponding to the binning for
  the spectrum in the top panel.  As is apparent from the position of
  the triangles in the confidence plot, the binning for the top
  spectrum is the compromise necessary to show SMM~J14011+0252's
  narrow line and SMM~J14009+0252's broader line, so this spectrum
  shows neither line detection to its best advantage.  In the
  confidence plot, the ordinate gives the number of nines (0.9, 0.99, etc.) in
  the confidence level; see text.
  \label{fig:mamo}}
  \vspace{2.5cm}
\end{figure*}

The upper panel of Figure~\ref{fig:mamo} is the full-band difference
spectrum between this pair of sources, showing detections for both: a
positive spike corresponds to an emission line from the first source
of the pair, SMM~J14011+0252 at $z \sim 2.55$, and a negative-going
spike corresponds to emission from the second source, SMM~J14009+0252
at $z \sim 2.95$. The binning in the top panel of
Figure~\ref{fig:mamo} does neither source justice because it is a
compromise between that best suited to the detection of J14011+0252's
narrow line and J14009+0252's broad line. 

\begin{figure*}
%\epsscale{.80}
\plotone{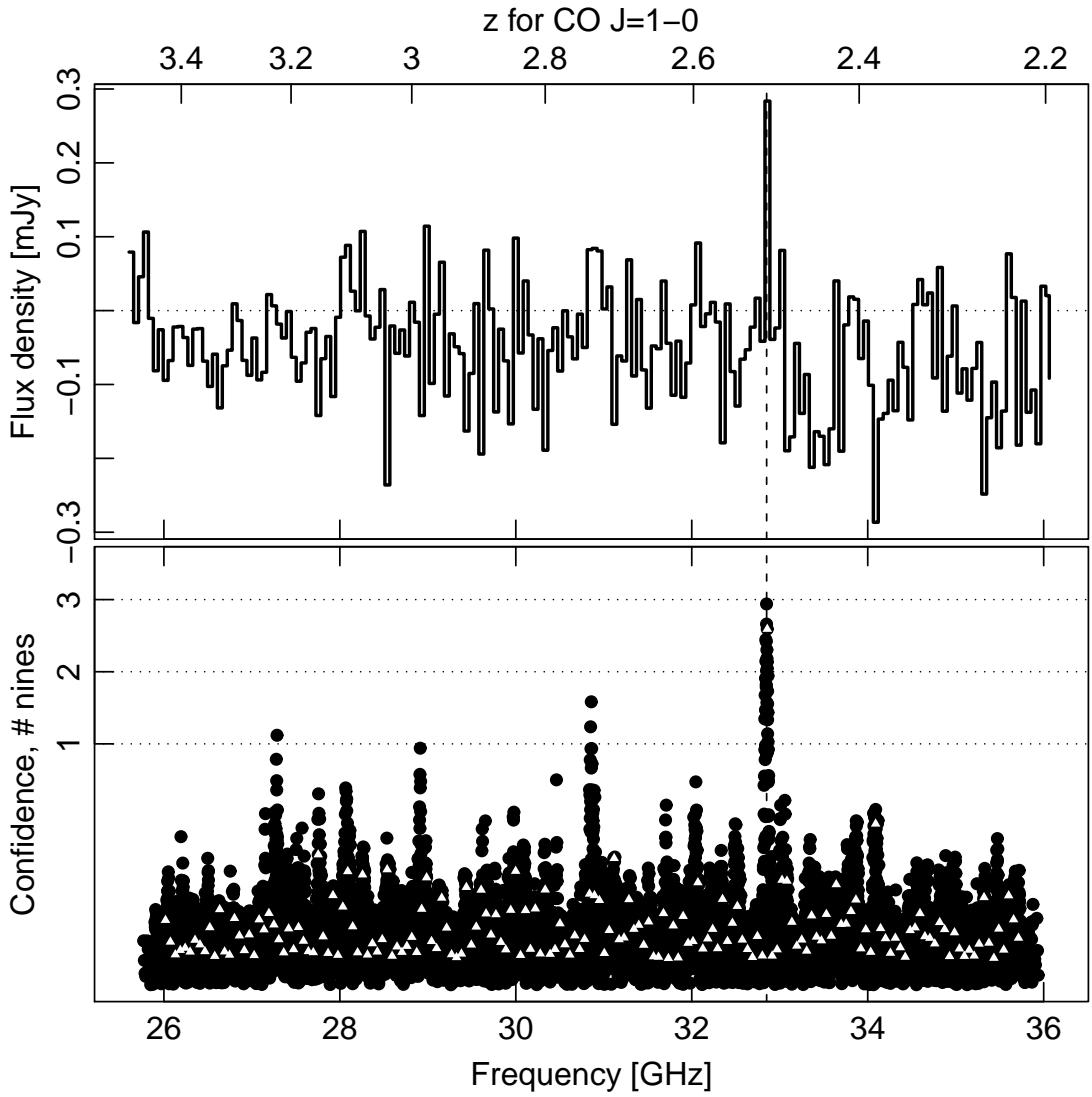} %fredPlot.eps}
\caption{Difference spectrum of SMM~J04431+0210 and SMM~J04433+0210 (upper
  panel) and confidence plot (lower panel).
  The triangles in the
  confidence plot mark the points corresponding to the binning for the
  spectrum in the
  top panel.
  See
  Fig.~\ref{fig:mamo}'s caption for an explanation of the confidence
  scale. \label{fig:fred}}
\end{figure*}

The lower panel in the figure contains a summary of many possible
binnings and shows the detections much more clearly.  We identify
lines by exploiting the fact that even broad extragalactic lines are
narrow on the scale of the Zpectrometer's bandwidth, searching for a
relatively narrow peak across a spectrum whose noise changes with
frequency.  This is a detection confidence plot, a statistically sound
quantitative embodiment of what an experienced observer would do by
eye: look to see whether a spectral channel or set of channels is
higher than its neighbors, within fluctuations.  Estimating noise
levels by eye is complicated by the changes in noise across the
spectrum's 25.6 to 36~GHz band; the fluctuation across all channels is
not a good measure of the fluctuations within individual channels.
Points in the figure summarize the results from our detection
statistic, which is described in the Appendix, over a wide range of
binning parameters.  Each dot in the confidence plot represents the
result from the detection statistic for one combination of bin width
and start channel index for the binning.  In Figure~\ref{fig:mamo}
(and Figure~\ref{fig:fred}), bin widths run from $n = 3$ to 10 and
starting bins run from 0 to $n-1$.  The triangles mark dots
corresponding to the bin width and starting bin for the spectrum in
the top panel, with one triangle at the center of each frequency bin.
Columns of dots are frequencies where the detection statistic is high
for a range of bin widths and center offsets.  Isolated dots are most
likely to be chance fluctuations emphasized by a particular set of
binning parameters.  Values on the vertical scale are the confidence
levels for line detection.  This is a very nonlinear scale, with units
in ``number of nines:'' i.e., levels of 0.9, 0.99, 0.999, etc.
Mathematically, the ``number of nines'' is equal to $ - \log_{10}(1-P)
- \log_{10}(N_{chan})$, where $P$ is the probability of detection and
the term with $N_{chan}$ corrects the scale to account for the
probability of a chance fluctuation given the number of channels in
the spectrum.  Without the channel number correction a ``$3\sigma$''
excursion would be unremarkable: if drawn from normally-distributed
noise, one should appear every 370 samples, on average, and each
confidence plot contains results from 52 different binnings, each with
50 to 200 channels, for a total of 9804 points.  Regions where the
columns coherently climb above 2 (0.99 confidence) give frequencies
where a potential detection is insensitive to exact binning,
indicating that the channel is reliably above or below the mean of its
neighbors.

\begin{figure*}
%\epsscale{.80}
\begin{minipage}[t]{3.3in}
\plotone{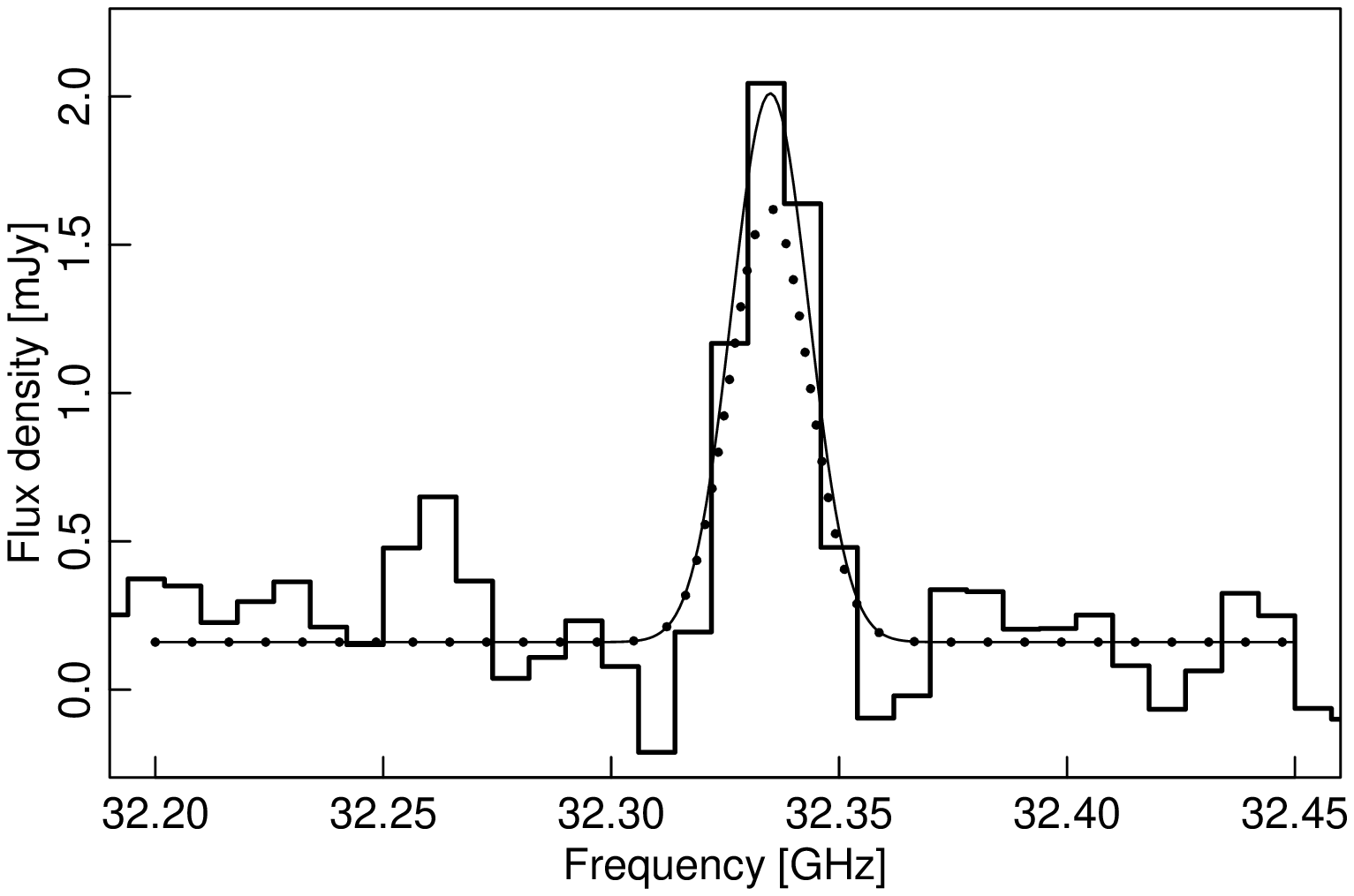} %prettyMax.eps}
\caption{Spectrum of SMM~J14011+0252 with Gaussian profile fit (curve) and
  Gaussian equivalent of the $3-2$ spectral fit with flux density
  divided by nine \citetext{dotted,
  \citealp{downes03}}. \label{fig:maxfit}}
\end{minipage}
\hfill
\begin{minipage}[t]{3.3in}
% trim to ylims 285 590
%\epsscale{.80}
\plotone{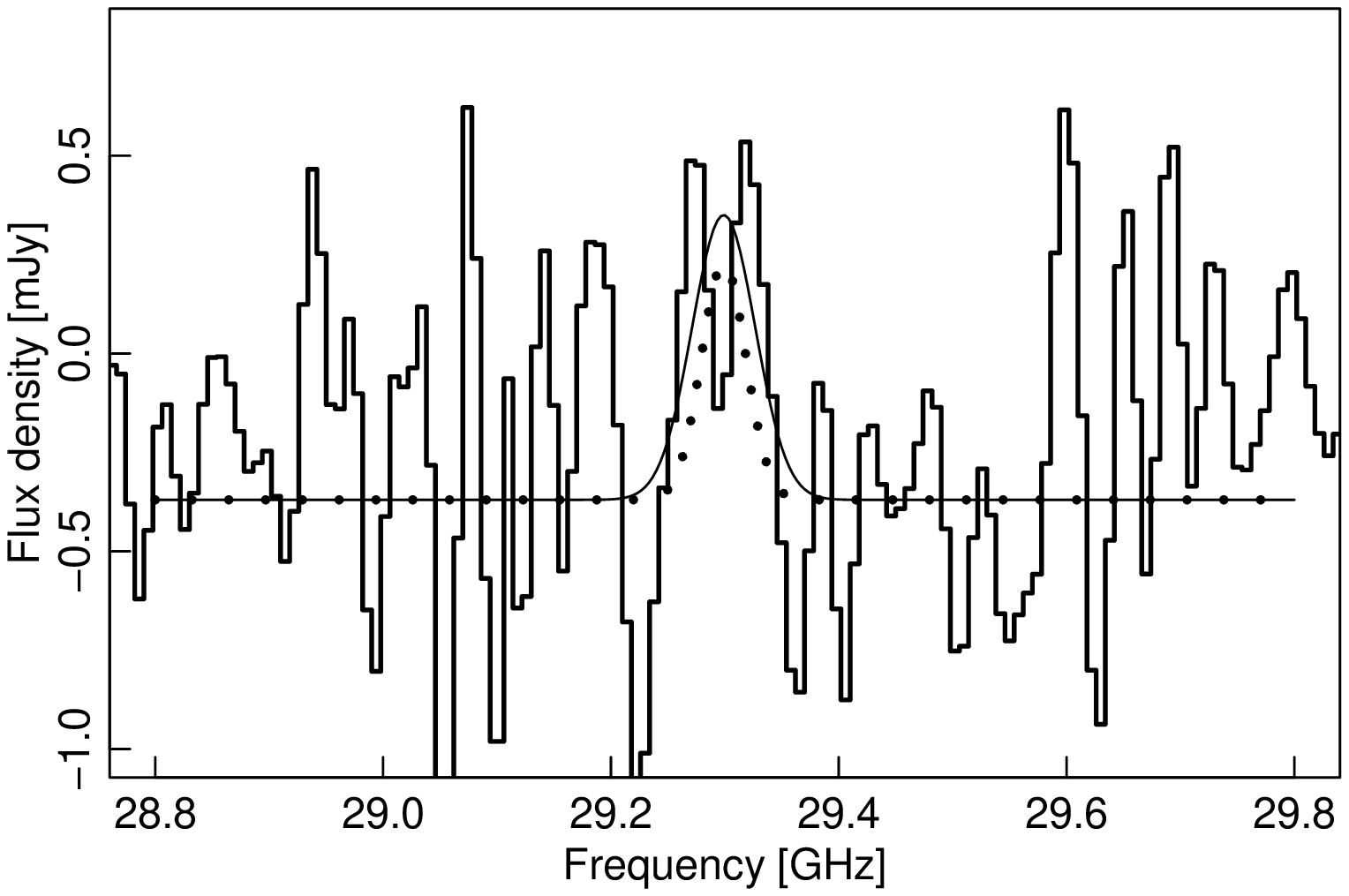} %sprettyMoritz.eps}
\caption{Spectrum of SMM~J14009+0252 with Gaussian profile fit (curve) and
  Gaussian equivalent of the $3-2$ spectral fit with flux density
  divided by nine \citetext{dotted,
  \citealp{weiss09}}. \label{fig:moritzfit}}
\end{minipage}
\end{figure*}

\begin{deluxetable*}{lcc}
\tabletypesize{\scriptsize}
%\rotate
\tablecaption{Summary of results and comparison to $3-2$ data from
  \citet{downes03} for SMM~J14011+0252. \label{tab:compMax}}
\tablewidth{0pt}
\tablehead{
 \colhead{Parameter} &
 \colhead{CO ($1-0$)} &
 \colhead{CO ($3-2$)}
}
\startdata
LSR Redshift & $2.5652 \pm 0.0002$  & $2.5652 \pm 0.0001$\\
Peak flux density, $S_{\nu}$ [mJy] & $1.85 \pm 0.20$ & $13.2 \pm 1$ \\
Line width [km~s$^{-1}$]           & $\leq208 $ & $190 \pm 11$ \\
Integrated line intensity [Jy\kms]      & $\sim 0.4 \pm 0.05$ & $2.8 \pm 0.3$ \\
Integrated intensity ratio, $I(3-2)/I(1-0)$ & $6.6 \pm 1.0$ & \ldots \\
Integrated brightness temperature ratio, $R_{3,1}$ & $0.76 \pm 0.12$ & \ldots \\
\enddata
\end{deluxetable*}

\begin{deluxetable*}{lcc}
\tabletypesize{\scriptsize}
%\rotate
\tablecaption{Summary of results and comparison to $3-2$ data from
  \citet{weiss09} for SMM~J14009+0252.  Structure in the spectral
  baseline rendered the line fits too poor
  for believable error estimates, so we quote the best fit without
  errors for line width and peak intensity.  Integrated intensity
  ratios and errors are derived from a bootstrap analysis; see text.
  \label{tab:compMoritz}}
\tablewidth{0pt}
\tablehead{
 \colhead{Parameter} &
 \colhead{CO ($1-0$)} &
 \colhead{CO ($3-2$)}
}
\startdata
LSR Redshift & $2.9346$  & $2.93450 \pm 0.00035$\\
Peak flux density, $S_{\nu}$ [mJy] &  $0.65$ & $5.4 \pm 0.9$ \\
Line width [km~s$^{-1}$]            & $643$ & $470 \pm 60$ \\
Integrated line intensity [Jy\kms]      & 0.44 & $2.7 \pm 0.3$ \\
Integrated intensity ratio, $I(3-2)/I(1-0)$ & $6.1 \pm 0.7$ & \ldots \\
Integrated brightness temperature ratio, $R_{3,1}$ & $0.67 \pm 0.08$ & \ldots \\
\enddata
\end{deluxetable*}

Vertical dashed lines through Figure~\ref{fig:mamo} show the
frequencies of the $1-0$ lines measured from fits to the
full-resolution unbinned $1-0$ spectra of the two sources shown over
small frequency ranges in Figures~\ref{fig:maxfit} and
\ref{fig:moritzfit}.  Since the $1-0$ and $3-2$ redshifts agree to
high precision, these also mark the $3-2$ redshifts.  An advantage of
observing with a fixed-tuned broadband system is that it is possible
to make direct comparisons of spectra and system performance for all
sources.  In all of the spectra we have taken with the Zpectrometer,
there have been no signs of spurious detections, and specifically none
at frequencies corresponding to either of these sources.

The smooth lines in Figures~\ref{fig:maxfit} and \ref{fig:moritzfit}
represent simple Gaussian profile interpretations of the $3-2$ line
parameters.  Tables~\ref{tab:compMax} and \ref{tab:compMoritz} give
the fit parameters and compare them with the $3-2$ parameters from the
literature. At the Zpectrometer's maximum frequency resolution,
spectral structure is correlated over about 3 bins; we correct for
that in linewidth measurements.

For SMM~J14011+0252, both the Figure and the tabulated data show that
the agreement between the $1-0$ and $3-2$ redshifts is excellent, and
that the linewidths are very similar \citetext{$3-2$ data from
  \citealp{downes03}, which agrees well with \citealp{frayer99}}.  The
integrated line intensity in Table~\ref{tab:compMax} is from the
Gaussian fit parameters to the line, with the $3-2$/$1-0$ integrated
intensity ratio equal to 6.8.  Individual velocity-integrated flux
densities (in this paper, the ``integrated'' in integrated intensities
and brightness temperatures implies integration over the full
linewidth in velocity) are computed from the fit parameters for a
Gaussian profile, $\int S_\nu \, {\rm d}v = 1.06 \, S_{\nu, peak} \,
\Delta v_{FWHM}$.

More revealing than the flux density ratio is the integrated $J = 1-0$
and $3-2$ brightness temperature ratio,
\begin{equation}
R_{3,1} = \frac{\int T(3-2) \, {\rm d}v}{\int T(1-0) \, {\rm d}v} =
  \frac{\int S_{\nu}(3-2) \, {\rm d}v}{\int S_{\nu}(1-0) \,
  {\rm d}v} \left( \frac{\nu_{1-0}}{\nu_{3-2}} \right)^2 \; ,
\end{equation}
where $S_\nu(x-y)$ is the flux density in the $J = x-y$ transition and
$\nu_{x-y}$ is the frequency of the transition.  Departures of this
ratio from unity reveal the departure from thick and thermalized
emission from gas in the Rayleigh-Jeans limit.  For this galaxy
$R_{3,1} = 0.76$, lower than the value of unity expected in this
single thick and thermalized limit.  Errors in Table~\ref{tab:compMax}
and following tables are at the 68.3\% ($1 \sigma$ equivalent for a
normal distribution) confidence level for the fits, as determined by
the R language's {\tt confint} routine.

SMM~J14009+0252's line is clearly detected when the binning
concentrates most of the flux into one channel (Fig.~\ref{fig:mamo}),
but nonideal structure in the baseline hinders precise line amplitude
and width measurements from full-resolution spectra.  Formal error
estimates are consequently not realistic.  Judging from the structure
in Fig.~\ref{fig:moritzfit}, the ripple in the spectrum causes the fit
to underestimate the true $1-0$ intensity and overestimate the
linewidth; an integrated intensity estimate from the fit parameters is
questionable.  Instead, we derive the integrated intensity and errors
from a bootstrap analysis of the numerator of
equation~(\ref{eq:zerotst}) for the binning that yields the highest
signal-to-noise detection, multiplying by a factor to account for flux
outside the rectangular bin for a Gaussian lineshape.  With $3-2$ data
from \citet{weiss09}, the $3-2$/$1-0$ intensity ratio is 6.1, or
$R_{3,1} = 0.67$.

\paragraph{SMM~J04431+0210 and SMM~J04433+0210}

\begin{figure}
%\epsscale{.80}
\plotone{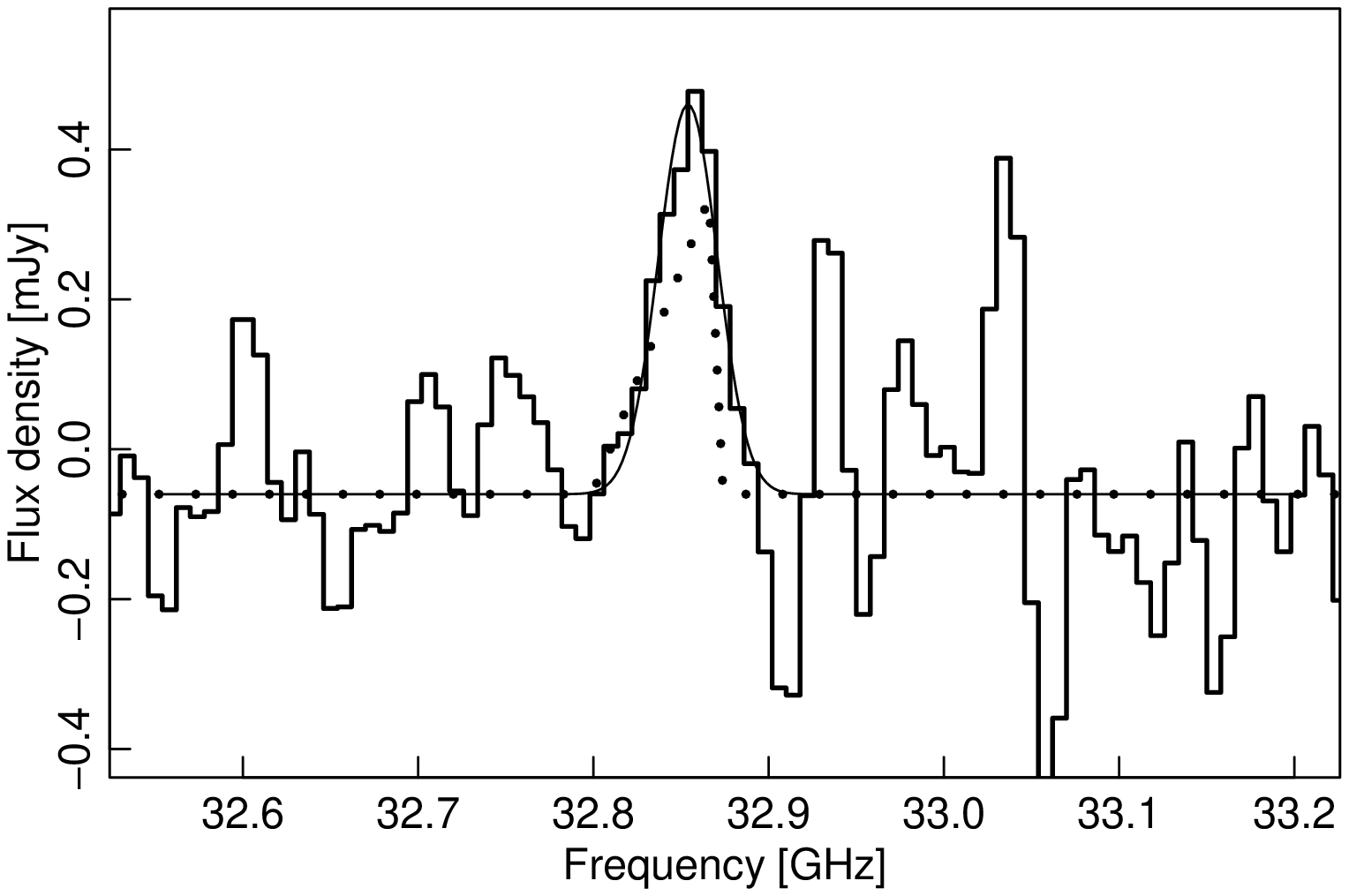} %prettyFredTri.eps}
\caption{Spectrum of SMM~J04431+0210 with Gaussian profile fit (curve) and
  triangular approximation to the $3-2$ spectral lineshape with flux density
  divided by nine \citetext{dotted,
  \citealp{neri03}}\label{fig:fredfit}}
\end{figure}

Figure~\ref{fig:fred} indicates that the only detection in this pair
is of SMM~J04431+0210; none of the negative-going structure is
detected with high significance.  The spectrum in the upper panel is
binned for the best signal to noise for this line.  The vertical
dashed line is the measured $1-0$ line frequency from a fit to the
full-resolution $1-0$ spectrum (Fig.~\ref{fig:fredfit}).  Since
agreement between the $1-0$ and $3-2$ redshifts are excellent, it
could equally well indicate the $3-2$ redshift.
Figure~\ref{fig:fredfit} is a full-resolution spectrum, with the solid
curve a single-component Gaussian profile fit.  The dotted line
represents the $3-2$ line's triangular shape \citep{neri03}, scaled to
the $1-0$ frequency, with the amplitude divided by nine.  Lineshape
differences between the two transitions are not large and could be due
to noise in either spectrum or to physical substructure with somewhat
different excitation conditions.  Table~\ref{tab:compFred} makes a
numerical comparison between the $1-0$ and $3-2$ \citep{neri03} line
parameters, with the $1-0$ parameters taken from the Gaussian profile
fit.  Taking the $3-2$ integrated intensity from \citet{neri03}, the
$3-2$/$1-0$ flux-density intensity ratio is 5.5 and $R_{3,1} = 0.61$.

The nondetection of SMM~J04433+0210 could be due either to a redshift
outside the Zpectrometer's coverage ($2.2 \leq z \leq 3.5$ for CO
$1-0$, $5.4 \leq z \leq 8.0$ for CO $2-1$) or to a low CO flux.  The
source has no known redshift, largely because its faint optical
counterpart is challenging for spectroscopy \citetext{$K = 22.4$:
  \citealp{frayer04}}, although the existence of a radio counterpart
suggests that it does not lie at a much higher redshift than the $z
\sim 2.5$ radio-preselected SMGs of the \citet{chap05} sample.  Given
that SMM~J04433+0210's 850~$\mu$m flux density of only $4.5 \pm
1.5$~mJy is barely above the detection threshold of the SLS
\citep{smail02}, within the range where noise fluctuations can
boost a weak detection significantly, we view low CO flux as the most
likely reason for its nondetection.

\begin{deluxetable*}{lcc}
\tabletypesize{\scriptsize}
%\rotate
\tablecaption{Summary of results and comparison to $3-2$ data from
  \citet{neri03} for SMM~J04431+0210. \label{tab:compFred}}
\tablewidth{0pt}
\tablehead{
 \colhead{Parameter} &
 \colhead{CO ($1-0$)} &
 \colhead{CO ($3-2$)}
}
\startdata
LSR Redshift & $2.5086 \pm 0.0007$  & $2.5094 \pm 0.0002$\\
Peak flux density, $S_{\nu}$ [mJy] & $0.58 \pm 0.08$  & $3.5$ \\
Line width [km~s$^{-1}$]           & $415 \pm 62$ & $350 \pm 60$ \\
Integrated line intensity [Jy\kms]      & $0.26 \pm 0.05$ & $1.4 \pm 0.2$ \\
Integrated intensity ratio, $I(3-2)/I(1-0)$ & $5.5 \pm 1.4 $ & \ldots \\
Integrated brightness temperature ratio, $R_{3,1}$  & $0.61 \pm 0.15$ & \ldots \\
\enddata
\end{deluxetable*}

\section{Discussion}\label{sec:disc}

We detected CO $J = 1-0$ emission from the three sources in our
samples that have known CO $3-2$ line parameters.  The similarities of
line redshifts and velocity widths for the $1-0$ and $3-2$ lines
justify a joint analysis of the two.  For the sample given in
Section~\ref{sec:obs}, the mean and standard deviation of the means is
$R_{3,1} = 0.68 \pm 0.08$.  Compared with other SMGs, the mean ratio
from our sample is slightly higher than the ratio $R_{3,1} = 0.48$ for
SMM J02399-0136 \citep{ivison10}, the brightest of the SLS galaxies in
850~$\mu$m continuum, and is comparable to the ratio for the highly
lensed SMG SMM~J2135--0102, $R_{3,1} = 0.68$ \citep{swinbank10,
  danielson10}.  For this sample of five SMGs with currently known
$1-0$ and $3-2$ fluxes, the ratio is $R_{3,1} = 0.64 \pm 0.10$.

A sample of five is not large, but neither is it so small that a few
additional observations can dramatically change the result.  For
example, if the ratio for most SMGs is really $R_{3,1} = 1$ and we
have been unlucky in our choice of galaxies, it will take observations
of another thirteen galaxies with unity ratio to reach a sample mean
ratio of $R_{3,1} = 0.9$.  Roughly speaking, observations of fifteen
additional galaxies will be needed to reduce the dispersion by a
factor of two.  We can also estimate the chances that we have drawn
five of five galaxies with $R_{3,1}$ below the true mean.  From the
binomial theorem, the probability is 0.03, for a symmetrical
distribution where each draw has a probability 0.5 of being above or
below the mean.  These simple considerations indicate that our
conclusion that $R_{3,1} \approx 0.6$ in SMGs is robust.  A recent
independent data set based on EVLA data gives $R_{3,1} \approx 0.55$
\citetext{\citealp{ivison10a}}, in agreement with
our value within errors.

The values of mean and dispersion in $R_{3,1}$ carry two linked
implications: the first one for mass estimates that rely on line
luminosity scaling, and the second for the state of the typical
interstellar medium in SMGs.

Observers have had to use incomplete data along with assumptions and
approximations to make mass estimates for distant galaxies.  In mass
derivations from millimeter wave spectroscopy of SMGs, which are most
often observed in the mid-$J$ lines, a frequent implicit assumption is
that the CO lines share the same excitation temperature $T_{ex}$ from
the observed mid-$J$ transition down to the $1-0$ line at the base of
the CO rotational ladder.  An alternative to the assumption of
constant $T_{ex}$ has been to fit observed line fluxes to a
single-component ISM radiative transfer model \citep[e.g.][]{weiss07}
and then use the model predictions for unobserved line fluxes.  The CO
$1-0$ data we report probe the ground-state emission of the CO
molecule to provide critical tests for these assumptions, which of
necessity invoke simple interstellar media.
  
The excitation temperature, $T_{ex}$, is a very general concept that
describes an energy density, whether kinetic, radiative, rotational,
vibrational, spin, etc. In observational molecular spectroscopy
$T_{ex}$ is the measured quantity; for the rotational transitions of
the CO molecule, the excitation temperature is the rotational
temperature $T_{rot}$.  In this paper we use the term thermalized in
its rigorous sense, to mean that $T_{rot}$ is equal for all
transitions of interest. Thermalized in this sense does not
necessarily imply that the molecule is in local thermodynamic
equilibrium (LTE), with rotational temperature equal to the kinetic
temperature of the surrounding \htwo\ molecules, although this is
often the case for low-$J$ CO transitions and is a common use of the
term.  In a rigorous context the term subthermal indicates that the
excitation temperature $T_{rot}$ for a given transition is below
$T_{rot}$ of a comparison transition from the same ensemble of
molecules, without reference to $T_{kin}$, $T_{rad}$, or any other
external bath.

\begin{figure}
% trim to ylims 285 590
%\epsscale{.80}
\plotone{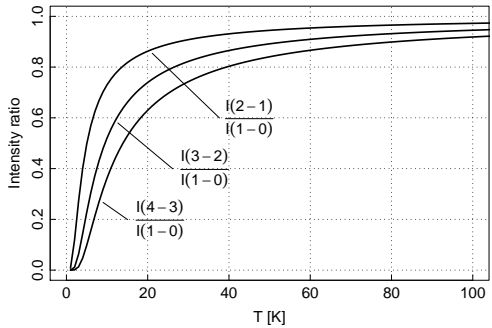} %planckRat3.eps}
\caption{Values of Planck function ratios for optically thick and
  thermalized gas for J $\le 4$.  The ratios reach the Rayleigh-Jeans
  limit, in which the radiation and kinetic temperatures are
  approximately equal, when the frequency $\nu$ of the higher
  frequency line is $h\nu/k \gg T$. \label{fig:planckRat}}
\end{figure}

Strictly speaking, the usual practice of extrapolating a constant
radiation temperature $T_{rad}$ measured in a mid-$J$ line for
lower-$J$ transitions is incorrect even if the density is high enough
that $T_{rot} = T_{kin}$: the kinetic temperature in SMGs is unlikely
to be high enough to drive $T_{rad}$ into the asymptotic
Rayleigh-Jeans limit.  As Figure~\ref{fig:planckRat} illustrates, the
Planck radiation (brightness) temperature ratios for even low-$J$
lines still climb toward unity for $T < 100$~K, so $I_\nu \propto
T_{rad} \neq T_{kin}$.  At a typical SMG dust temperature of about
40~K \citep{blain02} the correction from the Rayleigh-Jeans limit is a
factor 1.2 for the $4-3$ line, with corrections increasing with $J$.
If uncorrected, errors from the Rayleigh-Jeans approximation add
artificial rolloff to plots of CO line intensity vs.\ $J$ when those
plots involve scaling intensity by frequency squared to find the
collapse of excitation with $J$.

Returning to the observations, the small dispersion in $R_{3,1}$
justifies an empirical scaling prescription between the two
intensities for many sources, although there will doubtless be
excursions far from the mean for individual sources where scaling
fails.  In the local universe, an integrated brightness temperature of
the CO $1-0$ line, $I_{\rm CO} = \int T_{1-0}\,{\rm d}v$, has been
used extensively to estimate total molecular gas column densities.
The $1-0$ transition traces molecular gas with a wide range of
excitation and is relatively easy to observe in nearby sources.
Although optically thick, it can trace mass by ``cloud-counting'' many
individual clumps at different velocities within a telescope's beam
\citep[e.g.,][]{dickman86}.  Integrating $I_{\rm CO}$ over projected
area defines a luminosity $L^\prime_{\rm CO}$ that is directly
proportional to molecular gas mass; \citet{sdr92} derive a form
appropriate at cosmological distances as $L^{\prime}_{\rm CO} = I_{\rm
  CO}\,\Omega_s\,D^2_A$, where $\Omega_s$ is the source solid angle
and angular diameter distance $D_A$ accounts for cosmology.  $I_{\rm
  CO}$ is related to the \htwo\ column density $N$(\htwo) through the
$X$ factor, and $L^\prime_{\rm CO}$ to gas mass through the $\alpha$
factor.  $X$ and $\alpha$ differ by a constant factor and have been
calibrated for the $1-0$ line in Galactic, starburst, and ULIRG
environments; \citet{tacconi08} provide a brief critical review of
applying these scale factors in the high-redshift universe.

In the framework of scaling luminosity between different lines, the
mean value of $R_{3,1} = 0.64$ we find shows that a scaling factor of
about 1/0.64 = 1.5 is likely to be a more accurate predictor of the
$1-0$ integrated intensity from the $3-2$ integrated intensity than a
factor of unity.  To first order, this will increase gas masses
deduced from assuming $L^{\prime}_{\rm CO 1-0} = L^{\prime}_{\rm CO
  3-2}$ by a factor 1.5.

Departures from a ratio of unity are not limited to the $3-2$/$1-0$
pair.  Finding values other than unity from actual $1-0$ observations
of SMGs is becoming commonplace.  \citet{hainline06} and
\citet{carilli10} found similar behavior in their $1-0$ observations
of SMM~J13120+4242 and GN20, with an equivalent ratio $R_{4,1}$ of
0.26 (with some uncertainty from low-level emission in line wings) and
0.45, respectively.  The deviations from a simple thick and
thermalized model are in the same direction as we find, but are larger
and the scatter in the ratio is much higher.

\begin{figure}
% trim to ylims 285 590
%\epsscale{.80}
\plotone{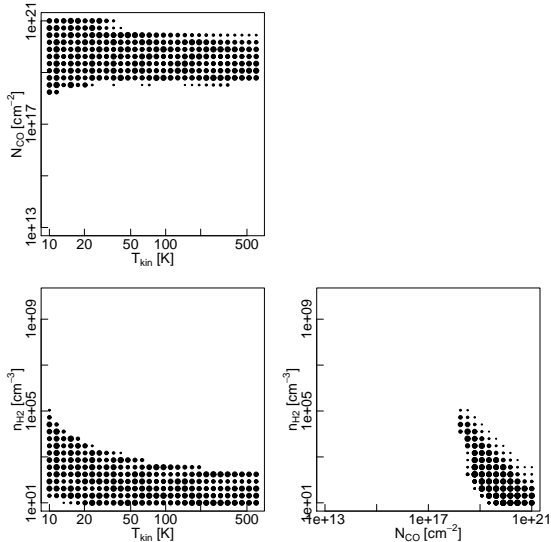} %radxfr_subtherm31_thick.eps}
\caption{Side views of the three-dimensional volume containing
  radiative transfer calculation results to identify the
  physical conditions that produce $R_{3,1} = 0.6 \pm 0.2$ by subthermal
  excitation.  The plot is a summary of 27k models filling the volume
  with equal spacing as seen in the projected views.  The points
  represent successful models, with size corresponding to the ratio:
  the largest diameter points are near the observed ratio of
  $R_{3,1} = 0.6$ and the smallest at $R_{3,1} = 0.4$ and 0.8.
  Results for optically thin
  lines have been removed from this summary to satisfy observational
  constraints.
  High-density, low-temperature results are not likely to be physical;
  see text.
  \label{fig:radxfr_subtherm}}
\end{figure}

Many of the earlier explanations for the lack of equality in line
brightness temperatures invoked subthermal (in the rigorous sense we
discussed) excitation of the mid-$J$ lines in such a simple,
single-component ISM as an explanation for decreasing $T_{ex}$ with
$J$.  We do not agree with this approach because it is an unphysical
distraction rather than a useful approximation.  Multi-line
observations of nearby galaxies (and, for that matter, Galactic giant
molecular clouds) show interstellar media with multiple components as
defined by lineshape, emission from different molecular species, or
detailed physical conditions.  If the $1-0$ and $3-2$ emission in the
SMGs we observe were from a single component of subthermally excited
gas, the $3-2$ intensity would be very sensitive to the detailed
physical conditions.  We modified an escape-probability radiative
transfer code \citetext{J.~Stutzki, priv.\ comm.; \citealp{stutzki85};
  cross-sections from \citealp{flo85}} to explore conditions matching
a conservatively broad range of $R_{3,1} = 0.6 \pm 0.2$.  To ensure
our model produced reasonable emissivities, we eliminated solutions
with optical depths less than unity in both lines.  The formal
solutions are in Figure~\ref{fig:radxfr_subtherm}.  Solutions with
densities much above 10$^3$\percc\ are not likely to be of physical
importance.  First, the high density solutions occur at low
temperatures, where the lines are extremely weak since they are close
to the $T = 9.6$~K background temperature at the model's $z = 2.5$.
Such lines will not dominate the molecular gas luminosity.  Further,
it is questionable whether the bulk of high density gas could be much
colder than the bulk of the dust, which has a temperature of about
$40$~K in SMGs \citep{blain02}.  Overall, the model indicates that
subthermally excited gas would have a density of a few$\times
10^2$\percc\ at a CO column density of $5 \times 10^{18}$~cm$^{-3}$ in
a 400\kms\ linewidth, with lower densities possible at higher columns.

Explanations for line ratios that require very specific physical
conditions (or a very constant gravitational lens amplification ratio
for the two lines) are not supported by the low dispersion of
$R_{3,1}$ from even our handful of SMGs.  It is unrealistic to expect
such a tight range of parameters from sample of galaxies selected
without regard for their CO emission properties, especially so when
observations of other lines directly contradict the special
conditions.  Interstellar media dominated by low-density material are
incompatible with detections of CO $J = 7-6$ from SMM~J14011+0252
\citep{downes03}, and $J = 5-4$ from SMM~J14009+0252 \citep{weiss09};
these lines are strong and rule out CO excitation that is collapsing
by $J = 3$.  The presence of excited gas in these sources does not
seem to be unusual: a compendium of CO excitation diagrams in
\citet{weiss07} shows that the brightest CO lines are in states $J$ of
5 to 6 for typical SMGs.

Brightness temperature ratios $R_{3,1} < 1$ are not peculiar to SMGs,
but are common in most galaxies.  \citet{yao03} and
\citet{mauersberger99} report mean values of $R_{3,1} = 0.66$ and
$R_{3,1} = 0.63$, respectively, in different samples of luminous
galaxies in the local universe.  \citet{iono09} find a mean $R_{3,1} =
0.48$ for dusty U/LIRGs at an altogether different redshift, and
\citet{aravena10} find $R_{3,1} = 0.61$ for a $z = 1.5$ BzK galaxy
selected by its rest-UV/optical colors.  At the same time,
\citet{bayet06} find that CO ladder excitation peaks at $J \sim 4$ for
the Galaxy and IC~342, and at $J \sim 7$ for starbursts, a type common
in the \citet{yao03} and \citet{mauersberger99} samples.  The
excitation rolloff seems to shift to much higher $J$ for very luminous
AGN hosts: \citet{vanderwerf10} find $J = 13-12$ is still strong in the
ULIRG Mrk~231, and \citet{bradford09} find strong lines to $J = 9$ in
the Cloverleaf QSO.

A pattern of apparently subthermal line ratios from an ISM that also
has strong emission from lines at higher $J$ is a clear signature of a
multi-component ISM.  In any given line from an ISM with a range of
physical conditions, the warmest optically thick regions with the
largest area filling factors will dominate the emission.  Basic
excitation considerations suggest strong $1-0$ emission from
distributed gas, while the $3-2$ emission is dominated by star-forming
and other regions with enhanced excitation.  High spatial resolution
observations of SMGs do show varying size with excitation.
\citet{danielson10} find systematic lineshape and size changes with
$J$ in CO emission toward one source. \citet{bothwell10} shows that
mid-$J$ CO emission regions' sizes are different than the sizes of the
star formation as traced by radio continuum.  Ivison
\citetext{\citealp{ivison10a}} make a comparison of emission sizes in
the CO $1-0$ and mid-$J$ transitions.  Evidence for higher excitation
in the centers of local U/LIRGs \citep{iono09}, which numerical models
imply may also characterize high-redshift SMGs \citep{narayanan09},
suggests that the balance between components is partly set on
galaxy-wide scales.  \citet{narayanan09} run hydrodynamic simulations
of galaxy mergers that yield SMG luminosities, finding that the
resulting ISM produces galaxy-averaged CO emission with line ratios
that mimic subthermal excitation from a single component ISM.  While
this qualitatively matches the data, the observed $3-2$/$1-0$ flux
ratio of $5.8 \pm 0.9$ is higher than the model results of $3 \pm 1$,
indicating that the gas is somewhat more excited than predicted.
Bringing the model results into better agreement with observations may
be useful in refining modeling prescriptions.  Overall, given the wide
range of galaxies that show a similar $R_{3,1}$ and lack of
correlation with most physical properties that \citet{yao03} find for
$R_{3,1}$, it is plausible that the different lines trace different
mixtures of regions in an ISM with a hierarchical (e.g. self-similar,
fractal) geometry, as is commonly found in high resolution
observations of our Galaxy and emerges from numerical simulations of
cloud structure.

At this stage we do not have enough lines to identify ISM structures
or to separate components through radiative transfer calculations, but
we can draw some simple conclusions from basic considerations.  The
brightnesses of the lines and correlation between $1-0$ and $3-2$
intensities indicate that the individual lines are optically
thick. Deviations from $R_{3,1} = 1$ then come from different
geometrical filling factors of related components in a multi-component
ISM, with a small contribution from failure to reach the asymptotic
Rayleigh-Jeans limit (Fig.~\ref{fig:planckRat}).

Finding a multi-component ISM in SMGs has implications for the choice
of relationship between CO $1-0$ intensity and mass, $M = \alpha
L^{\prime}_{CO}$.  The value used most often for SMGs is $\alpha =
0.8$~(M$_\odot$~km~s$^{-1}$~pc$^2)^{-1}$ \citetext{reviewed with
  caveats in \citealp{solomon05}}, which was derived for ULIRGs from
radiative transfer calculations based on an interpretation of a
$2-1$/$1-0$ integrated brightness temperature ratio below one as
evidence of subthermal excitation in ULIRGs \citep{downes98}.  An ISM
similar to other galaxies favors values for $\alpha$ closer to
calibrations from the dense parts of starburst nuclei or the Galaxy
\citetext{see \citealp{tacconi08} for a summary and references to
  individual studies}. Compounding the factor of
$\left(R_{3,1}\right)^{-1} = 1.5$, an increase in $\alpha$ by a factor
of about 2 would increase mass estimates for the SMGs by an overall
factor of about 3.  Increased mass estimates also imply decreased star
formation efficiency estimates.  Large masses derived from scaling
factors will run into upper limits set by dynamical mass measurements,
and may indicate that the local (Galactic) [CO]/[\htwo] abundance
ratio is lower than those in high-redshift SMGs.

In summary, we find observationally that a luminosity ratio of
$L^{\prime}_{CO}(3-2) \approx 0.6 \times L^{\prime}_{CO}(1-0)$ is more
appropriate than the customary assumption of equality.  Scaling the
$3-2$ integrated brightness temperatures to $1-0$ does seem to be
empirically justified, but a factor of $0.64^{-1} = 1.5$ is more
appropriate than unity.  Increasing the scaling factors for both line
integrated intensities and the conversion between line luminosity and
mass, as suggested by the similarity of $R_{3,1}$ in SMGs and local
luminous galaxies, would increase mass estimates from observed $3-2$
lines for SMGs further, unless conversion factors are allowed to
change with environment.  Increasing the scaling factor would increase
simple molecular mass estimates for SMGs and decrease their derived
star formation efficiencies.  It appears that simple line flux scaling
breaks down beyond $J = 3$, as although the trend in the line ratios
are similar, the dispersion in temperature ratios becomes
substantially larger, probably because the higher-$J$ lines probe past
the average peak excitation.  The large number of galaxies with
similar $R_{3,1}$ ratios, SMGs included, indicates that
single-component models are inadequate descriptions of what must be
more complex interstellar media.  Identifying galaxies with $R_{3,1}$
considerably different from the typical value of 0.6 will be valuable
in understanding the origin of the typical conditions in the ISM.

\acknowledgments This material is based on work supported by the
National Science Foundation under grant numbers AST-0503946 to the
University of Maryland and AST-0708653 to Rutgers University.  We
thank the NRAO staff, particularly P.\ Jewell, R.\ Prestage, K.\
O'Neil, R.\ Maddalena, B.\ Mason, A.\ Shelton, R.\ Norrod, and the GBT
software, mechanical design, and operations groups for their support
and contributions.  We thank J.\ Stutzki for a copy of his radiative
transfer code, A.\ Bolatto, D.\ Frayer, E.\ Ostriker, and R.\ Shetty
for perceptive comments, A.\ Wei\ss\ for confirming our first estimate
of J14009+0252's redshift before publication of his data, and
L.~Hainline for her contribution to the data reduction software and
for critical comments on the paper.  The National Radio Astronomy
Observatory is a facility of the National Science Foundation operated
under cooperative agreement by Associated Universities, Inc.

{\it Facilities:} \facility{NRAO (GBT)}.

\appendix

\section{Appendix: line detection confidence method}

Our detection confidence calculation is based on determining whether a
channel has a systematically different amplitude from its neighbors.
Setting a detection criterion for the line amplitude in a single
channel would be simple if we somehow knew the mean value with no line
present and had accurate knowledge of the system fluctuation within
the channel.  However, typical broadband spectra have offsets and other
large-scale structures in the spectral baseline, and the noise may
change with frequency as the receiver or system temperatures change.
Estimating noise parameters by calculating, for instance, the standard
deviation across frequency channels will be misleading in such cases.
Comparing the mean value of a channel with those of its neighbors is more
fruitful, as neighbors are likely to share offsets and noise.
Limiting the number of neighbors increases the uncertainty of noise
estimates in the spectral domain, however.  The time sequence of data
within each channel, sampled by the many subscans in a long
integration, provides information on the fluctuations in each
channel.

Our detection calculation is based on comparing mean values between
neighboring frequency channels, using noise estimates derived from the
time sequences of individual channels to calculate the statistical
significance of departures from the mean, with detection significance
framed in a classical hypothesis test.  Rather than testing for a
detection of a line with unknown amplitude, we find the probability
that the null case of no detection fails.  If a spectral line is {\it
  not} present in the parent data, then the mean value in some channel
$X$ will be equal to the mean value of its $M$ neighbors $Y_i$:
\begin{equation}
  \overline{X} - \frac{1}{M}\sum_{i=1}^{M} \overline{Y}_i = 0 \; ,
\end{equation}
within fluctuations from noise.

To estimate the fluctuation in each channel, we calculate sample
variances $S^2_X$ and $S^2_{Yi}$ in the time sequence of each channel,
derived from the $N$ 4-minute subscans in our final spectra.  This
approach provides estimates for individual channels, independent of
systematic structure across the spectrum.

Combining the difference in means and channel amplitude uncertainties,
a suitable test statistic is:
\begin{equation}
d = \frac{\left( \overline{X} - \displaystyle\frac{1}{M} \sum_{i=1}^{M}
                     \overline{Y}_i \right)}
                    {\sqrt{\displaystyle\frac{1}{NM}
                    \left( S^2_X + \displaystyle\sum_{i=1}^{M}
		    S^2_{Yi}
                    \right)} } \; .
\label{eq:zerotst}
\end{equation}
This is a useful form because it has a Student-$t$ distribution with
$(M+1)(N-1)$ degrees of freedom in the case that the signals in all
channels $X$ and $Y_i$ are from the same normally distributed parent
population. While this condition cannot be strictly true in general,
it can be a quick and reasonable approximation for channels close
together in frequency and over times short compared with atmospheric
changes; $t$-tests tend to be robust.  As values of $d$ become far
from zero, the value produced by a nondetection, it becomes less
likely that the true value is zero within fluctuations, and more
likely that a real deviation from a mean value of zero has been
detected.  For an emission line search, where even a large negative
excursion counts as a nondetection, the probability that the measured
weighted difference $d$ is consistent with zero is the one-tailed
test:
\begin{equation}
P({\rm fluctuation} \geq d) = P\left( d \geq t_{\alpha,(M+1)(N-1)}
\right)
= \int_d^{\infty}  T_{(M+1)(N-1)}(u) \:{\rm d}u
\; .
\label{eq:anaprob}
\end{equation}
In searching for positive and negative excursions, as we would with our
difference spectra, a two-tailed test is appropriate.

In practice, absolute probabilities given by
equation~(\ref{eq:anaprob}) are approximate because the assumptions of
stationary statistics with equal variances for all neighbors are not
strictly correct.  In addition, this analytical approach has no way to
accommodate weighting (e.g. for changes of atmospheric transmission
with time) during data analysis.  To counter these shortcomings, we
turned to the bootstrap technique \citetext{see \citealp{efron94}},
creating bootstrapped spectra by randomly drawing the the same number
of subscans as in the initial data set, with replacement (duplicates
are allowed), from the pool of subscans.  The bootstrapped spectra can
then be suitably weighted averages of individual subscans.  For the
examples in this paper, we made 2000 bootstrapped spectra for each
source pair, or 2000 examples of spectra we might have observed
assuming the pool of subscans is representative of all data (this is
the fundamental principle behind the bootstrap).  We then used
Eq.~(\ref{eq:zerotst}) as the ``studentized'' detection statistic,
since eliminating scale gives pivotal (variance stabilized) forms
with superior properties in bootstrap confidence calculations
\citep{efron94, zoubir04}.  The resulting distribution of $d$ should
be very closely normal, with the central limit theorem acting on sums
of the nearly-normal statistic $d$.  We verified the assumption by
comparing the distributions of $d$ and the normal distribution on
quartile-quartile plots, finding that the probabilities determined by
counting bootstrap results and by the normal approximation were the
same within fluctuations from finite sample length.

Counting bootstrapped results that satisfy a condition and then
normalizing by the number of bootstrap samples gives
distribution-independent probability estimates, with the probability
range bounded by the number of samples.  Samples of 2000 are suitable
for finding distribution-independent probabilities to about 0.99, for
instance.  To increase the confidence limit range we could have either
increased the number of bootstrap samples or used the empirical
agreement of the studentized $d$ to the normal distribution.  We chose
the latter as a conservative approximation: the observed bootstrap
result distribution should be, and empirically is, closely normal;
because a normal approximation is insensitive to a few extreme
results; and because of practical limits on computational time.  Even
with the calculation of $d$ in a compiled code function to speed
iterative calculations within the R language framework we use, typical
run times for a full range of binning were about a minute, so
increasing the number of samples by a few orders of magnitude from
2000 was impractical for interactive analysis.

\clearpage

%\bibliographystyle{apj} 
%\bibliography{ahtech,highz} 

\end{document}